\documentclass{article}
\usepackage{spconf}
\usepackage{color,xcolor}
\usepackage{epsfig}
\usepackage{graphicx}

\usepackage{array}
\usepackage{booktabs}
\usepackage{colortbl}
\usepackage{multirow}
\usepackage{float}
\usepackage{caption}
\usepackage[labelformat=simple]{subcaption}

\usepackage{footnote}
\makesavenoteenv{tabular}
\makesavenoteenv{table}

\usepackage{amsmath,amsfonts,amssymb,bm}
\usepackage[super]{nth}

\usepackage{changepage}
\usepackage{extramarks}
\usepackage{fancyhdr}
\usepackage{lastpage}
\usepackage{setspace}
\usepackage{soul}
\usepackage{xspace}

\usepackage{url}
\usepackage{hyperref}
\hypersetup{colorlinks=True, urlcolor=black}
\usepackage[numbers,sort&compress]{natbib}
\usepackage{algorithm, algorithmic}
\usepackage{enumitem}
\usepackage{verbatim}
\usepackage{pifont}
\usepackage[acronyms]{glossaries}
\glsdisablehyper

\newcommand{\sectdot}[1]{Sec.~\ref{sec:#1}}

\newcommand{\eqndot}[1]{Eqn.~(\ref{eqn:#1})}

\newcommand{\figdot}[1]{Fig.~\ref{fig:#1}}
\newcommand{\tbl}[1]{Table~\ref{tab:#1}}

\newcommand{\ignore}[1]{}

\makeatletter
\DeclareRobustCommand\onedot{\futurelet\@let@token\@onedot}
\def\@onedot{\ifx\@let@token.\else.\null\fi\xspace}

\makeatother

\definecolor{MyDarkBlue}{rgb}{0,0.08,1}
\definecolor{MyDarkGreen}{rgb}{0.02,0.6,0.02}
\definecolor{MyDarkRed}{rgb}{0.8,0.02,0.02}
\definecolor{MyDarkOrange}{rgb}{0.40,0.2,0.02}
\definecolor{MyPurple}{RGB}{111,0,255}
\definecolor{MyRed}{rgb}{1.0,0.0,0.0}
\definecolor{MyGold}{rgb}{0.75,0.6,0.12}
\definecolor{MyDarkgray}{rgb}{0.66, 0.66, 0.66}

\def\presec{\vspace{-0.2em}}
\def\postsec{\vspace{-0.2em}}

\title{Predicting Multi-Codebook Vector Quantization Indexes \\for Knowledge Distillation}

\name{Liyong Guo\textsuperscript{*,1}, Xiaoyu Yang\textsuperscript{*,1}, Quandong Wang\textsuperscript{1}, Yuxiang Kong\textsuperscript{1}, Zengwei Yao\textsuperscript{1}, \textit{Fan Cui}\textsuperscript{1},} \myname{\textit{Fangjun Kuang}\textsuperscript{1}, \textit{Wei Kang}\textsuperscript{1},  \textit{Long Lin}\textsuperscript{1}, \textit{Mingshuang Luo\textsuperscript{1}}, \textit{Piotr Żelasko\textsuperscript{2}}, \textit{Daniel Povey\textsuperscript{1}} \thanks{* stands for equal contribution}}

\address{\textsuperscript{1} Xiaomi Corp., Beijing, China \:\:\: \textsuperscript{2}Meaning.Team Inc, USA \\
\footnotesize{\texttt{\{guoliyong, xiaoyuyang6, dpovey\}@xiaomi.com, pzelasko@meaning.team}}}

\begin{document}
\ninept
\maketitle

\begin{abstract}
    Knowledge distillation (KD) is a common approach to improve model performance in automatic speech recognition (ASR), where a student model is trained to imitate the output behaviour of a teacher model. However, traditional KD methods suffer from teacher label storage issue, especially when the training corpora are large. Although on-the-fly teacher label generation tackles this issue, the training speed is significantly slower as the teacher model has to be evaluated every batch. In this paper, we reformulate the generation of teacher label as a codec problem. We propose a novel Multi-codebook Vector Quantization (MVQ) approach that compresses teacher embeddings to codebook indexes (CI). Based on this, a KD training framework (MVQ-KD) is proposed where a student model predicts the CI generated from the embeddings of a self-supervised pre-trained teacher model. Experiments on the LibriSpeech clean-100 hour show that MVQ-KD framework achieves comparable performance as traditional KD methods (l1, l2), while requiring 256 times less storage. When the full LibriSpeech dataset is used, MVQ-KD framework results in 13.8\% and 8.2\% relative word error rate reductions (WERRs) for non -streaming transducer on test-clean and test-other and 4.0\% and 4.9\% for streaming transducer. The implementation of this work is already released as a part of the open-source project icefall\footnote{https://github.com/k2-fsa/icefall}.
\end{abstract}

\presec
\section{Introduction}
\postsec
\label{sec:introduction}
 
 In the field of speech processing, significant improvements have been witnessed in self-supervised pre-training in recent years~\cite{baevski2020wav2vec,zhang2020pushing,hsu2021hubert,chen2022wavlm}. After pre-training on a very large amount of unlabeled data, the model is then fine-tuned with task-specific labeled data for downstream tasks such as ASR\cite{baevski2020wav2vec,hsu2021hubert,chen2022wavlm}, speaker verification\cite{wang2021fine, chen2022wavlm}, emotion recognition\cite{pepino2021emotion, wang2021fine}, etc.
 
To fully leverage the richness of unlabeled data, pre-trained models \cite{devlin2018bert, schneider2019wav2vec,hsu2021hubert,baevski2020wav2vec,chen2022wavlm} usually have a large number of parameters, ranging from hundreds of millions to several billions.. Although these models achieves state-of-the-art performance, they are impractical to be used in real-life scenarios due to their large model size and footprint. To deal with this, efforts have been made to utilize pre-trained models for improving smaller model's performance. Knowledge distillation (KD)\cite{hinton2015distilling}, also known as teacher-student training \cite{meng2019domain,manohar2018teacher,kurata2020knowledge,doutre2021improving} is applied to transfer information from a pre-trained teacher model to a typically-smaller student model, where the student model learns from labels generated from the teacher. Although the student model is typically of small size, there is an implicit problem being unnoticed: training efficiency. In traditional teacher-student training, the teacher labels are often float embeddings \cite{manohar2018teacher,bengio2012deep,kanda2018sequence} extracted on-the-fly, which would slow down the training if the teacher model is an extremely large pre-trained model. In addition to that, the maximum training batch size has to be reduced, leading to potential performance degradation. Otherwise, one could save the float type embeddings to disk before training and load them during KD training. However, the training speed would be constrained by the I/O and a huge amount of storage space is needed, making training impractical if the training corpus is large.

 Clustering or quantization is effective for representation learning \cite{hsu2021hubert, baevski2020wav2vec, baevski2019vq, bao2021beit}. Wav2vec2.0 \cite{baevski2020wav2vec} takes vector quantization (VQ) for clustering and the codebook vector is taken into computing contrastive loss. BEST-RQ \cite{chiu2022self} takes the VQ indexes as the pre-training labels.
 Inspired by these and borrowing ideas of residual vector quantization \cite{barnes1996advances} and the direct-sum codebooks \cite{barnes1993vector,barnes1995embedded}, we propose a trainable \textbf{M}ulti-codebook \textbf{V}ector \textbf{Q}uantization (MVQ) which compresses each embedding vector into a short sequence of 8-bit integer codebook indexes (CI)\footnote{https://github.com/k2-fsa/multi\_quantization}. 
 
Based on MVQ, a KD framework (MVQ-KD) is proposed by teaching a student model to predict CI generated from the embeddings at an intermeidate layer of a teacher model. This could solve the computation or storage issue of traditional KD methods\cite{yang2022knowledge,swaminathan2021codert}. For example, with the Hubert-Large model \cite{hsu2021hubert} whose dimension is 1024, it would cost 1976 gigabytes for storing 960 hours' of float-type teacher embeddings if 3-fold speed perturbation is used. However, only 7.72 gigabytes are needed for the corresponding CI in a 16-codebook MVQ setup, achieving a compression rate of 256. CI can be pre-computed and stored on disk at very low cost, which improves the training efficiency.

  
  The key experimental findings of this paper are:
  \vspace{-3pt}
  \begin{itemize}
    \item MVQ-KD achieves comparable performance as using traditional $l_1$ or $l_2$ losses, while saving 256 times storage space, avoiding the need of on-the-fly teacher label generation.
    \vspace{-3pt}
    \item The performance of MVQ-KD can be further improved with more codebooks.
    \vspace{-3pt}
    \item MVQ-KD is effective both for streaming and non-streaming transducer models.
    \vspace{-3pt}
  \end{itemize}

In the rest of this paper, \sectdot{mvq} illustrates the details of the MVQ algorithm. \sectdot{framework} briefly reviews the self-supervised pre-trained HuBERT model and presents the MVQ-KD framework. In \sectdot{experiments}, the experimental setup and results are described. Finally, conclusions are drawn in \sectdot{conclusions}.
\presec
\section{Trainable multi-codebook quantizer}
\postsec
\vspace{-0.1cm}
\label{sec:mvq}
 
 Consider a quantization module $\mathcal{Q}$ encoding vectors $\textbf{\textit x} \in \Re^D$ from a known distribution, into a fixed-size sequence of $N$ integers $0 \leq i_n < K$: let
 $\textbf{\textit i} = i_0,\ldots,i_{N{-}1}$, $\textbf{\textit i} \in {\{0,...,K{-}1\}}^N$. $\mathcal{Q}$ should have 
 \vspace{-0.4em}
 \begin{eqnarray}
          \textbf{\textit i} &=  \operatorname{Encode}(\textbf{\textit x}), \\
     \hat{\textbf{\textit x}} &=  \operatorname{Decode}(\textbf{\textit i}) .
\end{eqnarray}

We are interested in encodings that are as close as possible to optimal in an $l_2^2$-error
sense, i.e. that minimizes $E[||\hat{\textbf{\textit x}} - \textbf{\textit x}||_2^2]$.  To keep the encoding scheme
practical, we consider {\em direct-sum} codebooks, i.e. a scheme where the
$\operatorname{Decode}(\cdot)$ function sums over the codebooks:
\begin{equation}
 \operatorname{Decode}(\textbf{\textit i}) = \sum_{n=0}^{N-1} {\textbf{\textit c}}_{i_n}^{(n)} .
\end{equation}
This requires $K\times N$ codebook centers $c_k^{(n)} \in \Re^D$. The length of the integer sequence $N$ can be referred to as the number of codebooks.

When encoding using direct-sum codebooks, it is impractical to enumerate all possible encodings as the encoding space is $\mathcal{O}(K^N)$. One straightforward way is to choose the codebook center sequentially that reduces the residual error the most while keeping other codebook centers unchanged. This heuristic is not guaranteed to yield the lowest residual error as the codebook centers from different codebooks are not jointly evaluated. Therefore, an iterative encoding scheme is proposed to improve the aforementioned heuristic, which efficiently searches for better encodings.

\vspace{-0.5em}
\presec
\subsection{Iterative encoding scheme}
\label{sec:encoding}
\postsec

Assuming the number of codebooks to be $N$, $\operatorname{Encode}(\cdot)$ compresses a float vector $\textit x$ to $N$ CI. The encoding function has $N$ independent neural classifiers, which generate an intial estimate of CI. The encoding process $\operatorname{Encode}(\cdot)$ is implemented as follows :
\begin{itemize}
    \item Choose the initial codebook entries $\textbf{\textit i}$ as the arg-max of $N$ independent logistic regression classifiers.
     
    \item For e.g. 3 iterations, refine the codebook entries: $\textbf{\textit i} \leftarrow \operatorname{Refine}(\textbf{\textit x},\textbf{\textit i})$, each time using the refined index generated from the previous call of $\operatorname{Refine}(\cdot,\cdot)$.
    
    \item Return the refined indexes. They will be used as input for $\operatorname{Decode()}$ and as label to train the classifiers (see \sectdot{MVQ training})
    
\end{itemize}

 The function $\operatorname{Refine}(\textbf{\textit x}, \textbf{\textit i})$ is the most essential part in $\operatorname{Encode}(\cdot)$. It is given initial codebook indexes $\textbf{\textit i} = i_n, 0 \leq n < N$, and returns possibly-improved codebook entries $\hat{\textbf{\textit i}} = \hat{i}_n, 0 \leq n < N$:
\begin{itemize}
  \item For each codebook $0 \leq n < N$ and each index $0 \leq k < K$, compute modified residual $\hat{\textbf{\textit x}}- \textbf{\textit x}$ assuming we let the $n$'th codebook center be $k$ but leaving all
     the other codebooks with their initial values (the ones in $\textbf{\textit i}$).
  \item For each codebook $n$, sort the residuals above and store indexes of the $J$ smallest residuals. 
  \item Construct a sub-problem that has $N/2$ codebooks, with each codebook being of size
    $J^2$, by summing pairs of $J$-best codebook centers, combining $n=0$ with $n=1$,
    $n=2$ with $n=3$, and so on.
  \item Recurse, call $\operatorname{Refine}(\cdot, \cdot)$ on the smaller
    sub-problem.
  \item The result $\hat{\textbf{\textit i}}$ can be computed from the answer to the sub-problem and the indexes of the $J$-best entries for each codebook.
 \end{itemize}
 The recursion terminates when the sub-problem only has one codebook and the index resulting in the lowest residual is selected. The refined index $\hat{\textbf{\textit i}}$ can be returned recursively.

\vspace{-1.0em}
\presec
\subsection{Training and Inference Procedure}
\label{sec:MVQ training}
\postsec
The trainable parameters in the quantization module are codebook centers $c_k^{(n)}$ and $N$ logistic-regression classifiers $\mathcal{C}_n$. For each float vector $\textbf{\textit x}$ and its encodings $\operatorname{Encoder}(\textbf{\textit x})=\textbf{\textit i}$, the training loss $\mathcal{L}$ consists of two parts:
\vspace{-0.2em}
 \begin{align}
     {\mathcal L} &= {\mathcal L}_{\mathrm{residual}} + {\mathcal L}_{\mathrm{prediction}}, \\
           &= ||\textbf{\textit x} - \operatorname{Decode}(\textbf{\textit i})||_2^2 + \sum_{n=1}^{N}  -\log \mathcal{C}_n(\textbf{\textit x})_{\textit{i}_{n}}
 \vspace{-0.4em}
 \end{align}
where $\mathcal{C}_n(\textbf{\textit x})_{\textit{i}_{n}}$ is the probability of predicting $\textit{i}_{n}$ in $\mathcal{C}_n$. The first term ${\mathcal L}_{\mathrm{residual}}$ is the reconstruction loss, i.e $l_2^2$ residual and optimizes $\mathcal{C}_n$. The second term ${\mathcal L}_{\mathrm{prediction}}$ is the prediction loss and encourages the neural classifiers to select the encoded indexes $\textbf{\textit i}$ improved by $\operatorname{Refine}(\cdot, \cdot)$. By doing so, the initial estimate given by $\mathcal{C}_n$ is expected to be close to the refined CI. Training is performed based on gradient descent with Adam~\cite{2014Adam} optimizer. During inference, the encoding process described in \sectdot{encoding} is repeated to generate CI for test data.
\vspace{-0.9em}
\presec
\subsection{Analysis of reconstruction loss}
\postsec
 
Rate-distortion theory can be used to evaluate the reconstruction performance of the proposed quantization algorithm. For quantizers with various numbers of codebooks trained on HuBERT-L\cite{hsu2021hubert} with embedding dimension of 1024, the first column of \tbl{shannon} shows the relative reconstruction loss (RRL), defined as the mean of the squared reconstruction error $||\hat{\textbf{\textit x}} - {\textbf{\textit x}}||_2^2$ divided by the mean of $||{\textbf{\textit x}} - {\mu}_{x} ||_2^2$, i.e. the average sum-square of {\textbf{\textit x}} after mean normalization. The last column shows the best possible distortion assuming the 1024 dimensions were normally and independently distributed. This comes from the rate-distortion equation for a memoryless Gaussian source~\cite{thomas2006elements}: $R(D) = \frac{1}{2}\log_2(\sigma_x^2 /D)$, with $\sigma_x^2 = 1$, the bit-rate per dimension $R$ set to $\frac{8N}{1024}$ since there are $N$ codebooks of 8 bits each, and solving for distortion $D$. Although this is a lower bound on the distortion, the values are actually higher than the second column as the HuBERT-L embeddings are not strictly independent and Gaussian.

When applied to features that are normally distributed and independent, the algorithm achieves RRL (see second column in \tbl{shannon}) that is within 10\% of the Shannon lower bound, so the performance in terms of reconstruction loss does not have that much potential for further improvement as far as our purposes are concerned.

\begin{table}[!ht]
\vspace{-0.4cm}

    \centering
    \captionsetup{position=below}
    \caption{Relative Reconstruction Loss (RRL) and Shannon distortion if embedding dimensions were i.i.d. normal}
    \label{tab:shannon}
    \begin{tabular}{c c c  c }
        \toprule

        N & RRL(HuBERT) & RRL(Gaussian) & Shannon distortion \\
        \midrule 
        1 & 0.517 &  0.992 & 0.989 \\ 
        4 & 0.356 &  0.969 & 0.958 \\ 
        8 & 0.278 & 0.938 & 0.917 \\ 
        16 & 0.225 & 0.876 & 0.841 \\ 
        32 & 0.206 & 0.760 & 0.707 \\ 
        \bottomrule
    \end{tabular}
    \vspace{-0.5cm}
    
\end{table}
\presec
\section{Proposed distillation framework}
\postsec
\label{sec:framework}

\subsection{Self-supervised pre-trained HuBERT}

Recently, self-supervised pre-training has shown promising results \cite{baevski2020wav2vec, hsu2021hubert} in ASR. Among these methods, HuBERT\cite{hsu2021hubert} is one of the most effective frameworks. HuBERT model comprises of three parts: convolutional neural network (CNN) encoder, transformer and acoustic unit discovery system. The CNN encoder processes raw speech waveform $\textit{\textbf{w}}$ and generates embedding $\textit{\textbf{X}}=\textit{\textbf{x}}_{1:T}$. The acoustic unit discovery system then produces the hidden unit target $z_t$ for each $\textit{\textbf{x}}_t$ using k-means clustering. Before feeding embedding $\textit{\textbf{X}}$ to the transformer to generate contextualized representations, a set of randomly selected timestamps are masked. The self-supervised pre-training objective is to predict the correct hidden unit $z_t$ for both masked and unmasked timestamps with $\mathcal{L} = \alpha \mathcal{L}_m + (1-\alpha) \mathcal{L}_u$,
where $\mathcal{L}_m$ and $\mathcal{L}_u$ is the CrossEntropy loss for masked and unmasked timestamps and $\alpha$ is a tunable coefficient. During training, $z_t$ is refined to improve the clustering quality. After pre-training, HuBERT can be fine-tuned with labeled speech for ASR tasks\cite{hsu2021hubert,wang2021fine}. 


\vspace{-0.8em}
\presec
\subsection{Traditional KD Methods for Neural Transducers}
\postsec
Neural transducers is a powerful modelling framwork E2E ASR. It has gained more popularity recently due to its natural support for streaming and superior performance. To further improve the performance of neural transducer, knowledge distillation (KD), or teacher-student training, is common used. During KD, a student model is trained to imitate the output of a teacher model. Depending on the teacher's output, different loss functions can be applied for KD training. Kullback-Leibler (KL) divergence is commonly used if teacher labels are distributions whereas $l_1$ or $l_2$ are more appropriate for continuous feature. As a transducer generates a 3-D distribution lattice, directly applying KL-divergence is computational intractable. \cite{panchapagesan2021efficient} used a collapsed version of the distribution lattice to reduce computation, whereas \cite{yang2022knowledge} approximated the distribution lattice with its one-best alignment. Both methods\cite{panchapagesan2021efficient,yang2022knowledge} pre-computed the teacher labels and stored them to disk, which could be problematic for large training corpora. Instead of utilizing the output distribution, embedding features are another straightforward teacher label. \cite{swaminathan2021codert} uses the $l_2$ loss between the encoder embeddings of teacher and student model for KD. Let teacher embedding $\textbf{TE}^{l_{th}}= \textbf{TE}^{l_{th}}_1,...,\textbf{TE}^{l_{th}}_T$ be the embedding extracted from the $l_{th}$-th layer of the teacher model and $\text{SE}^{l_{st}} =\textbf{SE}^{l_{st}}_1,...,\textbf{SE}^{l_{st}}_T$ be the embedding at $l_{st}$-th layer of the student model, the KD loss function is:
\begin{align}
    \mathcal{L}_{embedding} = \sum_{t=1}^{T} \textbf{Dist} (\text{TE}^{l_{th}}_t, \textbf{LossNet}(\text{SE}^{l_{st}}_t)),
\end{align}
where $\textbf{Dist}$ is any function (e.g $l_1$, $l_2$) that measures the distance between two vectors and $\textbf{LossNet}$ is usually a linear layer that maps $\textbf{SE}^{l_{st}}$ to the same dimension as $\textbf{TE}^{l_{th}}$. $\textbf{TE}^{l_{th}}$ are generated on-the-fly since teacher and student are jointly trained\cite{swaminathan2021codert}. However, this will inevitably affect batch size or training speech, which could affect the performance of the student model. 

\vspace{-0.5em}
\presec
\subsection{MVQ-based KD for Neural Transducers}
\postsec

\begin{figure}[t]
  \centering
  \includegraphics[width=\linewidth]{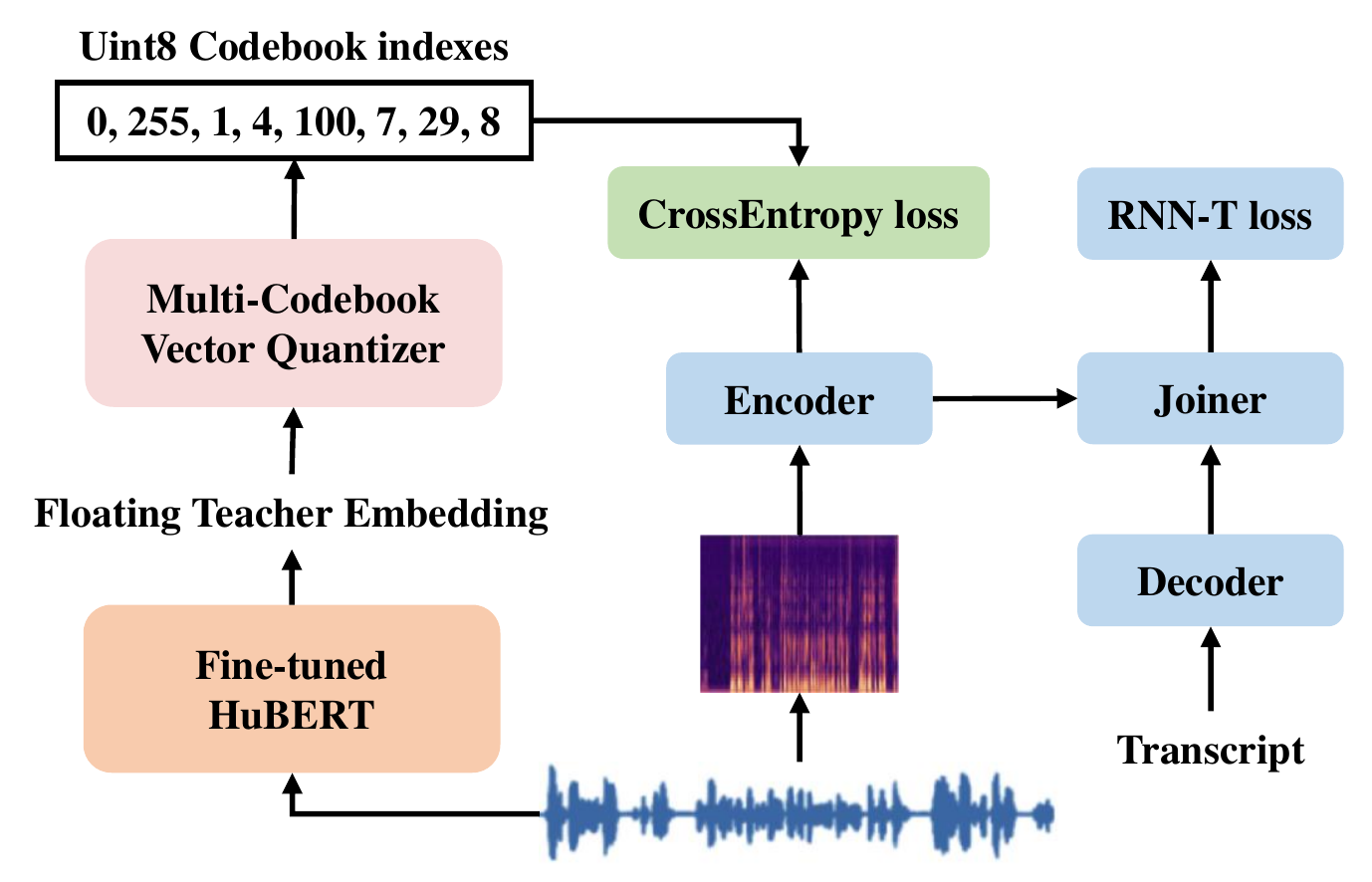}
  \caption{MVQ based teacher-student learning.}
  \label{fig:multi_task_leanring}
   \vspace{-0.4cm}
\end{figure}

To alleviate the aforementioned issues, we propose to apply MVQ on the embedding extracted from an intermediate layer of teacher $\mathbf{TE}^{l_{th}}_t$ and compress it to CI. Then, instead of regressing $\mathbf{TE}^{l_{th}}_t$ using $l_1$ or $l_2$ loss, the student model is trained to predict its CI. Let $N$ be the number of codebooks, MVQ compresses $\mathbf{TE}^{l_{th}}_t$ to $\textbf{\textit{i}}_t =(i_{t,1},...,i_{t,N})$, with $i_{t,n}$ representing which entry in the $n$-th codebook is chosen at $t$-th frame, the loss function is:
\vspace{-0.5em}
\begin{align}
    \mathcal{L}_{cb} = \sum_{t=1}^{T} \sum_{n=1}^{\text{N}} \textbf{CrossEntropy}(\mathbf{i}_t, \textbf{LossNet}(\text{SE}^{l_{st}}_t)),
    \label{eqn:cb loss}
    \vspace{-0.1cm}
\end{align}
where $\textbf{LossNet}$ is a module consisting of a linear layer with softmax activation that transforms the student embedding to probabilities. At each timestamp $t$, MVQ-KD performs $N$ independent classification. If the number of codebook centers in each codebook is 256, each entry in $\textbf{\textit{i}}_t$ can be represented by an 8-bit integer. Therefore, they can be pre-computed and store on disk at very low cost. With $N$=16 and teacher embeddings of 1024 dimensional, MVQ achieves a compression ratio of 256, significantly increasing KD training's scalability compared to traditional KD methods. \figdot{multi_task_leanring} illustrated the proposed KD framework for MVQ-based KD. 

Different from non-streaming transducer models, streaming transducer model only has limited access to future context and tends to emit symbols later. Therefore, applying \eqndot{cb loss} directly on a streaming transducer could be problematic as it may force the student to guess into the future. Inspired by\cite{yang2022knowledge}, a time-shift variable $\delta$ is introduced to address the temporal mismatch between teacher and student model. This leads to a modified version of $\mathcal{L}_{cb}$:
\vspace{-0.5em}
\begin{align}
    \mathcal{L}_{cb} = \sum_{t=1}^{T-\delta} \sum_{n=1}^{\text{N}} \textbf{CrossEntropy}(\mathbf{i}_t, \textbf{LossNet}(\text{SE}^{l_{st}}_{t+\delta})),
    \label{eqn:cb loss streaming}
\end{align}

The codebook loss $\mathcal{L}_{cb}$  will be used as an auxiliary loss to the original transducer loss:
\begin{align}
    \mathcal{L}_{total} = \mathcal{L}_{transducer} + \lambda \mathcal{L}_{cb},
\end{align}
where $\lambda$ is a tunable scale of the auxiliary loss.


\presec
\section{Experiments}
\postsec
\label{sec:experiments}

\subsection{Datasets and Model}
The LibriSpeech ASR corpus \cite{panayotov2015librispeech} was used for all experiments. The full dataset contains 960h hours of transcribed audio. Among these, the ``train clean 10'' subset was used for comparison with other baseline models and hyper-parameter tuning. During training, SpecAug\cite{park2019specaugment} and speed perturbation with rate 0.9 and 1.1 are used for data-augmentation. MUSAN\cite{musan2015} is used for noise-augmentation. The output vocabulary has 500 subword units and WERs are reported on test-clean and test-other sets using beam search. 

The large version of HuBERT\cite{hsu2021hubert} is adopted to initialise the encoder of the teacher transducer model and finetuned on full LibriSpeech. The student model is also a transducer model with a reworked version of Conformer\footnote{https://github.com/k2-fsa/icefall/tree/master/egs/librispeech/ASR}\cite{gulati2020conformer} as encoder. In streaming experiments, we apply causal convolutions and blockwise-limited right context in attention and train the student model with dynamic chunk size \cite{zhang2020unified}. Pruned RNNT loss\cite{kuang2022pruned} is used for computing $\mathcal{L}_{transducer}$. Details of teacher and student model are listed in \tbl{model details}. The embeddings of the 18-th transformer block are extracted for CI generation as we believe this layer contains rich information while being less difficult to learn from. Consequently, KD is carried out on the 9-th layer of the student model to share the same relative position as in the teacher model. A randomly sampled subset of 1000 audios are used to generate embeddings and train the quantizer $\mathcal{Q}$. Then, CI are generated by feeding the whole training set to $\mathcal{Q}$ and stored to disk. To compare with $l_1$ and $l_2$ losses, the 18-th layer's float embeddings are also stored. 

\begin{table}[ht]
\vspace{-0.2cm}
    \centering
    \setlength{\abovecaptionskip}{0pt}%
    \caption{Details of teacher and student models}
    \begin{tabular}{ccc}
    \toprule
         & Teacher model & Student model\\
    \midrule
    Encoder & HuBERT-Large & Conformer \\
    Encoder dim & 1024 & 512 \\
    Encoder layer & 24 & 12\\
    Num Params & 318M & 78M \\
    \bottomrule
    \end{tabular}
    
    \label{tab:model details}
\vspace{-0.4cm}
\end{table}

\vspace{-0.1cm}
\presec
\subsection{Impact of the Number of Codebooks}
\postsec

Table \ref{tab:num_codebook} demonstrates the impact of codebook numbers ($N$) on the MVQ distillation, from 1 to 32. With $N$=1, the quantization degenerates into the traditional VQ. For different $N$, the scale of $\mathcal{L}_{cb}$ is tuned individually for optimal performance. Using only one codebook, the student model already outperforms the baseline model. The student model consistently improves as $N$ increases. This also accords with the fact that MVQ achieves lower reconstruction error if more codebooks are used (see \tbl{shannon}). As $N$=16 achieves similar WERs compared to $N$=32 while doubling the compression rate, $N$=16 is selected in future experiments.

\begin{table}[!h]
\vspace{-0.5cm}
\captionsetup{position=below}
\caption{WERs with different number of codebooks}
\label{tab:num_codebook}
    \centering
    \begin{tabular}{c c c c}
    \hline
    \toprule
        $N$ & compression-rate & test-clean & test-other \\ 
        \midrule
        baseline, 0 & - & 6.83 & 18.19 \\ 
        1 & 4096& 5.67 & 15.77 \\ 
        2 & 2048& 5.58 & 15.27 \\
        4 & 1024& 5.39 & 14.68 \\ 
        8 & 512& 5.14 & 14.51 \\ 
        16 & 256 & 5.01 & 13.80 \\ 
        32 & 128 & \bf 4.99 & \bf 13.68 \\ 
        \bottomrule
    \end{tabular}
    \vspace{-0.0cm}

    \vspace{-8pt}
    
\end{table}




\presec
\subsection{Teacher-student learning with different losses}
\postsec

Table ~\ref{tab:base_l1_l2_cb16} compares the WERs of student model trained using MVQ KD with other traditional $l_1$ and $l_2$ losses. To ensure the fairness of comparison, on-the-fly teacher label generation is not adopted for $l_1$ and $l_2$ as this will limit the batch size. Experiments are only carried out on clean-100 subset as storing teacher labels of 960h audio for $l_1$ and $l_2$ loss computation is impractical. The scales of all auxiliary losses are tuned individually and only the setup with the lowest WERs are reported. The following observations can be made from \tbl{base_l1_l2_cb16}. First, the proposed KD framework successfully improves the performance of the student model. Both MVQ KD and traditional $l_1$ and $l_2$ method are able to reduce the WERs of the student model, indicating the effectiveness of using an intermediate layer for KD. Second, although KD with $l_2$ loss results in the lowest WERs, MVQ still achieves comparable performance, while being able to be flexibly applied in larger scale of experiment. 
During the experiments, it is found that the embedding values of HuBERT model are unstable, sometimes ranging from -2000 to +3000. Applying $l_1$ and $l_2$ loss requires special design such as clamping the embedding values, whereas MVQ-KD is less sensitive to this.

\begin{table}[ht]
\vspace{-0.5cm}
\captionsetup{position=below}
\caption{WER for baseline and  distillation with different losses}
\label{tab:base_l1_l2_cb16}
    \centering
    \begin{tabular}{l c c}
    \toprule
        config  &test-clean & test-other \\
        \midrule
        baseline & 6.83 & 18.19 \\ 
        $l_1$  & 5.1  & 13.69 \\ 
        $l^2_2$  & 4.99  & 13.39 \\
        MVQ, $N$=16  & 5.01 & 13.80 \\
        MVQ, $N$=32  & 4.99 & 13.68 \\ 
        \bottomrule
    \end{tabular}
    \vspace{-0.5cm}
    
\end{table}

\subsection{Training with full LibriSpeech}

To further demonstrate the effectiveness and robustness of MVQ, experiments are scaled up to the full LibriSpeech for both non-streaming and streaming student transducer models. For non-streaming models, relative WER reductions (WERRs) of 13.8\% and 8.2\% are achieved on test-clean and test-other. For streaming models, both WERs and latency are shown for different $\delta$. The latency is measured against the word-level alignment obtained from a hidden Markov model using \cite{mcauliffe2017montreal}. To get a reasonable estimate for $\delta$, the locations of posterior peaks in the lattice of teacher and student models are compared. The following three key observations can be made. First, MVQ-KD improves the accuracy of streaming model if a sensible $\delta$ is selected. The model trained with $\delta=0$ has higher WERs than the baseline model, while the model trained with larger $\delta$ outperforms the baseline model, achieving WERRs of 4.0\% and 4.9\% with $\delta=5$. Second, MVQ-KD reduces the latency of streaming models. Setting $\delta$ to 4 or 5 not only improves the model accuracy, but also encourages the model to emit faster, achieving a latency reduction of 0.1 seconds compared to the baseline model. Third, as $\delta$ increases, the latency also increases while the WER decrease, suggesting that $\delta$ controls the trade-off between model latency and model accuracy.



\begin{table}[ht]
\vspace{-0.5cm}
    \centering
    \captionsetup{position=below}
    \caption{WER of models trained with full LibriSpeech}
    \label{tab:full_libri}
    \begin{tabular} { l c c c}
    \toprule
         &test-clean& test-other & latency (s) \\ 
        \midrule
        \multicolumn{3}{l}{\textbf{Reference models}}\\
        Teacher, HuBERT-L & 1.9 & 3.94 & - \\
        Baseline, non-streaming & 2.69 & 6.11 & - \\ 
        Baseline, streaming & 3.03 & 7.98 & 0.335\\
        \midrule
        \multicolumn{4}{l}{\textbf{MVQ-KD trained model}}\\
        Non-streaming & \textbf{2.32} & \textbf{5.61} & - \\ 
        Streaming, $\delta=0$ & 3.13 & 7.9 & 0.165 \\
        Streaming, $\delta=4$ & 2.99 & 7.64 & \textbf{0.235} \\
        Streaming, $\delta=5$ & \textbf{2.91} & \textbf{7.59} & 0.259 \\
        \bottomrule
    \end{tabular}
    \vspace{-0.1cm}
    
\end{table}
\vspace{-1.4em}

\presec
\section{Conclusions}
\postsec
\label{sec:conclusions}
\vspace{-0.1cm}

In this paper, we present an efficient and effective knowledge distillation (KD) framework for neural transducers based on a novel Multi-codebook Vector Quantization (MVQ) algorithm. With a fine-tuned self-supervised pre-trained model, we show that our framework achieves comparable performance as the traditional $l_1$ and $l_2$ losses, while being much faster or requiring hundreds of times less storage. We also demonstrate that the proposed KD framework is effective both for non-streaming and streaming student model. In future works, we would like to incorporate multiple teacher layers for KD to further improve the student model. Since MVQ is a general quantization algorithm, we would also like to explore the feasibility of applying MVQ-KD on other speech processing tasks. 


\newpage
\section{References}
\vspace{-0.5em}
\begingroup
\renewcommand{\section}[2]{}
\bibliographystyle{IEEEbib}
\bibliography{refs}
\endgroup
\end{document}